\documentclass[sigconf]{acmart}

\usepackage{caption,booktabs}
\usepackage{multirow}
\usepackage{balance}

\begin{document}
\settopmatter{printacmref=false} % Removes citation information below abstract
\renewcommand\footnotetextcopyrightpermission[1]{} % removes footnote with conference information in first column
\pagestyle{plain} % removes running headers

%
% The "title" command has an optional parameter, allowing the author to define a "short title" to be used in page headers.
\title{Causal Inference in Higher Education: Building Better Curriculums}

\author{Prableen Kaur}
\affiliation{%
  \institution{University of Minnesota, Twin Cities}
  \city{Minneapolis}
  \state{Minnesota}
}
\email{kaur0016@umn.edu}

\author{Agoritsa Polyzou}
\affiliation{%
  \institution{University of Minnesota, Twin Cities}
  \city{Minneapolis}
  \state{Minnesota}
}
\email{polyz001@umn.edu}

\author{George Karypis}
\affiliation{%
  \institution{University of Minnesota, Twin Cities}
  \city{Minneapolis}
  \state{Minnesota}
}
\email{karypis@umn.edu}

%
% The abstract is a short summary of the work to be presented in the article.
\begin{abstract}
Higher educational institutions constantly look for ways to meet students' needs and support them through graduation. Recent work in the field of learning analytics have developed methods for grade prediction and course recommendations. Although these methods work well, they often fail to discover causal relationships between courses, which may not be evident through correlation-based methods. In this work, we aim at understanding the causal relationships between courses to aid universities in designing better academic pathways for students and to help them make better choices. Our methodology employs methods of causal inference to study these relationships using historical student performance data. We make use of a doubly-robust method of matching and regression in order to obtain the casual relationship between a pair of courses. The results were validated by the existing prerequisite structure and by cross-validation of the regression model. Further, our approach was also tested for robustness and sensitivity to certain hyper parameters. This methodology shows promising results and is a step forward towards building better academic pathways for students.
\end{abstract}

\keywords{Causal Inference; Learning Analytics; Average Treatment Effect; Matching.}

\maketitle 
\section{Introduction}
It is a known fact that many students struggle with choosing courses that align with their career goals. Their decisions rely on advice from their peers, their academic advisors and other resources. As universities constantly try to to help the students in making better choices throughout their degrees, designing better tools and resources has become an active area of research. Recent methods of course recommendations employ several approaches to provide an academic path to the students. However, these methods recommend courses that are correlated to each other and may sometimes fail to identify courses that are not correlated but have a causal relationship. This follows from the fact that "correlation does not imply causation".

In this study, we focus on studying the casual relationships between pairs of courses in order to answer the following question: "If a student takes course $X$, will it cause them to do well in a subsequent course $Y$?". To answer this, we have used methods of causal inference for observational data. The methods developed for causal inference focus on obtaining causal relationships in experimental setups. However, studying causal inference with observational data is challenging due to the lack of control over the environment variables involved. Another major challenge is the difficulty in validating both the methodology and results due to the absence of ground truth information.  

We use matching methods and a regression-based approach in order to overcome this problem in observational data. This ensures that our methodology is doubly robust. We evaluated both the accuracy and the robustness of our methodology. The results were validated using the existing prerequisite structure and it was observed that our methods identify additional courses that do not appear in the prerequisite structure. Thus, our methodology may be used to modify/aid in building better academic pathways to guide students and help them achieve their goals.

\section{Methodology overview}
In order to obtain the causal relationships between pairs of courses, we make use of methods of causal inference for observational data. The following subsections discuss our approach in a detailed manner.

\subsection{Casual Effect and the Average Treatment Effect (ATE)}

Rubin \cite{Rubin} defines a causal effect as the difference in the outcome $y$ in the presence of a treatment $T$ versus the absence of the treatment $T$. This can be formulated as: 
\begin{equation}
  y_i(T=1,Z) - y_i(T=0,Z),
\end{equation}
where $Z$ includes any other features that influence the outcome $y$. Notice that this method of defining the causal effect of treatment $T$ assumes that we can observe a particular individual simultaneously under both conditions, i.e., with and without the treatment. However, it is impossible to observe the outcomes of a individual simultaneously with and without treatment, one of the potential outcomes will always be missing. This observation is described as the "fundamental problem of causal inference" \cite{Justin}. This is shown in Figure 1, where we are trying to find the causal effect of taking the course CSCI 2021 (Computer Architecture) on the grade of the student in the subsequent course CSCI 4041 (Algorithms and Data Structures). We would need for student A to both take and not take the course CSCI 2021 at the same time, and observe the grade that he/she took in course CSCI 4041.
\begin{figure}[t]
  \centering
  \includegraphics[width=0.9\linewidth]{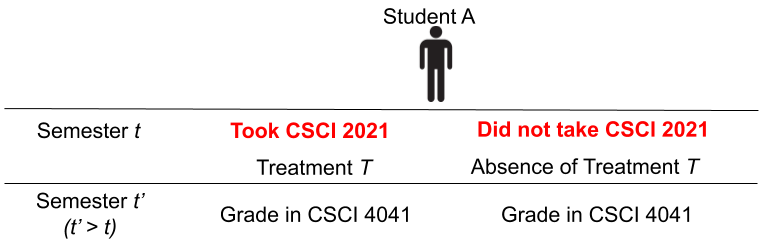}
  \caption{Fundamental problem of Causal Inference.}
\end{figure}

We can overcome this problem by observing a group of individuals that are on average the same except for the presence or absence of the treatment $T$. Once we have these groups, we can then apply the above idea, to obtain the Average Treatment Effect (ATE): 
\begin{equation}
    ATE = E(y|T=1) - E(y|T=0).
\end{equation}
Equation 2 holds if the following two assumptions are satisfied \cite{Justin}: First, the \textit{Stable Unit Value Treatment Assumption}, which states that the outcome $y$ is independent of both the mechanism by which the treatment $T$ is assigned to an individual and the treatment $T$ that is assigned to other individuals in the set-up. Second, the \textit{strong ignorability} assumption which states that an individual's assignment to a treatment condition is not a function of that individual's potential outcomes. 

For the purpose of finding the causal relationships between pairs of courses, we define the treatment $T$ as a prior course $X$ (taken in semester $t$) and the outcome $y$ as the received grade in a subsequent course $Y$ (taken in semester $t' > t$). The ATE will then give us the causal effect of taking course $X$ on the subsequent course $Y$; thus, answering our research question.

\subsection{Matching}
We need to find a Treatment group $\mathbb{T}$ and Control group $\mathbb{C}$ respectively, such that, the individuals from both groups are on average the same except for the presence and absence of the treatment. For our study, we defined the groups as follows (shown in Figure 2):
\begin{enumerate}
    \item \underline{Treatment group $\mathbb{T}$}: All students that have taken prior course $X$ and scored above a C grade.
    \item \underline{Control group $\mathbb{C}$}: All students that have not taken prior course $X$ or have taken prior course $X$ but scored a C grade or below. 
\end{enumerate}

We now need to ensure that individuals in the two groups are on average similar to each other to accurately compute the ATE. This is done by performing 1:1 matching between the two groups based on the assumption that students can be identified as similar based on: total credits (this shows their progress in the program), GPA (this shows their performance in the program), and prior courses (these can be used to identify similar students based upon student history). Here, both total credits and GPA are computed up to and not including the semester in which the course $Y$ was taken. Based on this, the following distance metric was used to compute the distance between two individuals $(s^T, s^C)$ from each group respectively:   
\begin{displaymath}
    Dist = \sqrt{(\textrm{GPA diff})^2 + (\textrm{Credits diff})^2 + (\textrm{1 - Jaccard Sim})^2},
\end{displaymath}
where, \begin{displaymath}
    \textrm{GPA diff} = \textrm{GPA}(s^T) - \textrm{GPA}(s^C),
\end{displaymath}
\begin{displaymath}
    \textrm{Credits diff} = \textrm{Credits}(s^T) - \textrm{Credits}(s^C),
\end{displaymath}
\begin{displaymath}
  \textrm{Jaccard Sim} = \frac{|A \cap B|}{|A \cup B|}.
\end{displaymath}
Here, $A$ and $B$ are the lists of courses taken prior to course $X$ by the two individuals $s^T$ and $s^C$,  respectively.
\begin{figure}[t]
  \centering
  \includegraphics[width=0.9\linewidth, keepaspectratio]{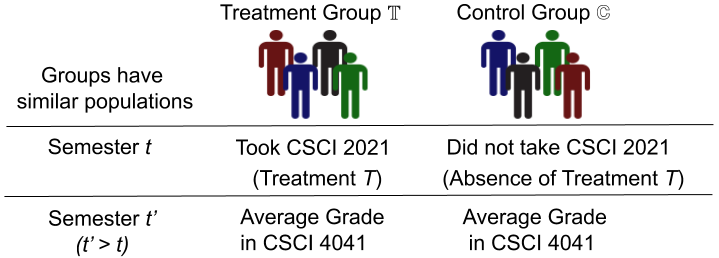}
  \caption{Defining Treatment and Control groups based on prior courses and performance.}
\end{figure}
We now perform 1:1 matching using a greedy approach. First, we find the group with the minimum number of students, i.e. min (|$\mathbb{T}$|, |$\mathbb{C}$|). Then, for every student in this group, we find the student belonging to the other group that has the lowest distance from it (i.e. the most similar student). These two students form a 1:1 matching pair. We then eliminate these two matched students from further matches. In this way, we find a pair for every student in the smaller group until the distance between two students exceeds some cut-off distance (e.g., 0.5) or every student from the smaller group has found a matching pair, whichever occurs first. (Note that in order to ensure consistency, the distances computed are scaled between 0-1).

\subsection{Computing ATE}

Now that we defined the Treatment and Control groups, we can finally compute the ATE. We compute the ATE using two methods:
\begin{enumerate}
    \item \underline{Based on means $(ATE_{means})$}: Difference in the average outcome of the treatment and control group respectively, i.e.,
    \begin{equation}
        ATE_{means} = E({y_T}|Z) - E({y_C}|Z),
    \end{equation}
    where $Z$ includes any other features that influence the outcome $y$.
    \item \underline{Based on Regression $(ATE_{reg})$}: We model the students' grade in course $Y$ using a binary indicator variable $T'$ (indicates the presence/absence of the prior course $X$) and the remaining features i.e., GPA, total credits and binary indicator variables for prior courses. We then use the coefficient of the binary indicator variable $T'$ as a measure of ATE. Mathematically, the regression equation can be written as:     
    \begin{equation}
        y_i = \beta_0 + \beta_{ATE}T' + \sum_{k=1}^{P}{\beta_kZ_k} + \epsilon_i,
    \end{equation}
    where $T'$ is a binary variable that indicates if the student is in the Treatment group $\mathbb{T}$ or not, and $Z_k$  indicates all the other covariates: GPA, total credits and binary indicators for all prior courses of course $Y$.
\end{enumerate}

Computing ATE using these two approaches is a method of being doubly robust. This ensures that if either of these two approaches, i.e., matching or regression, accurately estimates the ATE, we obtain an accurate estimate of ATE \cite{Justin}.

%Results Table
\begin{table*}
\begin{center}
\caption{Results of our method on Cohort 1 and Cohort 2 (* indicates a statistically significant difference in means with significance level = 0.01).}
  %\label{tab:commands}
  \begin{tabular}[th]{cccccccc}
    \toprule
    {} & {} & \multicolumn{3}{c}{Spring 2002 -- Spring 2010} & \multicolumn{3}{c}{Fall 2010 -- Spring 2016}\\
    \hline
    Course $Y$& Course $X$ & \multirow{2}{4em}{$ATE_{means}$} & \multirow{2}{6em}{$ATE_{reg}$} & \multirow{2}{4em}{\textrm{RMSE}}& \multirow{2}{4em}{$ATE_{means}$} & \multirow{2}{6em}{$ATE_{reg}$}& \multirow{2}{4em}{\textrm{RMSE}}\\
    Sem t' & Sem t & {}& {}& {}& {}& {}& {}\\ 
    \midrule
    \texttt {\multirow{2}{6em}{CSCI 4041 Algo \& DS}} & CSCI 1933 & 0.19* & 0.23 &0.18 &0.05 & 0.04 & 0.22\\
    \texttt {}&CSCI 2021&-0.06*&-0.05&0.18&0.07*&0.09&0.21\\
    \hline
    \texttt {\multirow{3}{6em}{CSCI 5103 Op. Systems}} & CSCI 1133 & 0.07 & 0.11 &0.23& 0.22* & 0.16&0.19\\
    \texttt {}&CSCI 1933 & -0.05 &-0.06 & 0.21 &0.24* & 0.24 &0.23\\
    \texttt {}&CSCI 2021&0.03 & 0.06 & 0.2 &0.19*&0.13&0.21\\
    \bottomrule
  \end{tabular}
\end{center}
\end{table*}

\section{Experimental Design}
\subsection{Dataset}
For the purpose of this study, we used a dataset from the Computer Science department at the University of Minnesota. The data spans around 14 years and consists of transcript-like information of all students over these years. The general scheme followed is that students enrolled at the university have to select and register for courses every semester. Students are awarded credits for every course they take based on an A--F grading scale. Further, the university has a general requirement that courses in which a student receives less than a C-- do not count toward satisfying degree requirements. 

Following are some additional constraints we applied on the dataset: of all students enrolled, we only consider students that actually received their degree. Further, we removed courses that belonged to other departments, non-academic courses (like independent/directed study or field study) and instances that did not receive a letter grade in the A--F grading scale. We did not consider offerings in the summer semester and finally, we only retained valid students that had at least two consecutive semesters with valid courses.

As our dataset consists of data ranging from the year 2002 -- 2016, we split the data into two cohorts: Spring 2002 -- Spring 2010 and Fall 2010 -- Spring 2016. This was done to take into account the change in the degree prerequisite structure of the undergraduate Computer Science program at the University of Minnesota after Spring 2010.

We define a valid pair of courses as a combination of course $Y$ taken in semester $t'$ and it's prior course $X$ taken in semester $t (t' > t)$. Here, courses $X$ and $Y$ are chosen based on the following criteria, choose course $Y$ as any course taken after semester 1 (so that a prior course exists), that has at least 10\% students who scored below a C grade, that has at least 100 students that have taken the course. For every valid course $Y$, choose course $X$ as any course that students have taken before taking the course $Y$ and that has a minimum support of 100 students.

\subsection{Evaluation methodology and metrics}
Evaluation in the field of causal inference is in general difficult. We evaluate our methodology using two ways: First, we compare our results to existing prerequisite courses and their relationships to other courses in order to validate our results. This validates that our method is capable of identifying these courses as causal.

Second, we evaluate the outcome of our regression model from Eq. (4) through a k-fold cross validation and evaluate the performance through RMSE between the predicted and true outcome grade in course $Y$ using the formula:
\begin{displaymath}
  \textrm{RMSE} = \sqrt{\sum_{i=0}^{n}{(\hat{y_i} - y_i)^2}},
\end{displaymath}
where $n$ is the number of records in the test set.

Finally, we report the average $\beta_{ATE}$ obtained over the k-folds as the Average Treatment Effect for each pair of courses. We also compute the standard deviation of $\beta_{ATE}$ among each k-fold evaluation. These evaluation methods ensure that our regression model represents the true population and accurately models the grades for students in the course $Y$. 

\section{Results}
Upon applying these methods to our dataset, we computed the ATE between multiple pairs of valid courses using the two methods on the two cohorts as described above.

%Sensitivity Table
\begin{table*}
\begin{center}
  \caption{Sensitivity Analysis: $Sim(d,d')$ for every pair of distances $d$ and $d'$.}
  %\label{tab:commands}
  \begin{tabular}[t]{ccccccc|ccccccc}
    \toprule
    {} & \multicolumn{6}{c}{Spring 2002 -- Spring 2010} & \multicolumn{6}{c}{Fall 2010 -- Spring 2016}\\
    {}& 0.1&0.3&0.4&0.5&0.6&0.9& 0.1&0.3&0.4&0.5&0.6&0.9\\
    \midrule
    \texttt 0.1&*&0.556&0.514&0.514&0.542&0.569&*&0.667&0.762&0.762&0.714&0.714\\
    \texttt 0.3&0.556&*&0.962&0.808&0.782&0.833&0.667&*&0.900&0.867&0.933&0.900\\
    \texttt 0.4&0.514&0.962&*&0.846&0.821&0.872&0.762&0.900&*&0.967&0.967&0.933\\
    \texttt 0.5&0.514&0.808&0.846&*&0.974&0.923&0.762&0.867&0.967&*&0.933&0.967\\
    \texttt 0.6&0.542&0.782&0.821&0.974&*&0.949&0.714&0.933&0.967&0.933&*&0.967\\
    \texttt 0.9&0.569&0.833&0.872&0.923&0.949&*&0.714&0.900&0.933&0.967&0.967&*\\
    \bottomrule
  \end{tabular}
\end{center}
\end{table*}

\subsection{Causal Results}
We focus on the results of a few courses as shown in Table 1. For every pair of courses $(Y,X)$, we have computed $ATE_{means}$ and the $ATE_{reg}$ obtained over all the k-folds for both cohorts. Looking at the results, for CSCI 4041 (Algorithms and Data Structures) we find that lower level courses, like CSCI 1933 (Intro to Algorithms and Data Structures), show a positive causal effect i.e., taking CSCI 1933 causes students to perform better in CSCI 4041. This is further validated by the fact that CSCI 1933 is an existing prerequisite for CSCI 4041. We also observe that the course CSCI 2021 (Computer Architecture) has a negative causal effect in the earlier cohort as compared to the positive causal effect in the latter cohort. This may be due to changes in the course structure from the earlier cohort to the latter i.e., CSCI 2021 has shifted from a more theoretical based syllabus to a more practical syllabus which would then lead to a better understanding of the material for CSCI 4041. 

Next, if we look at the higher level course such as CSCI 5103 (Operating Systems), we find that the lower level prerequisite courses CSCI 1133 and CSCI 2021 are found to have a positive causal relationship for both cohorts. However, we also find that CSCI 1933 has reversed relationships from the first cohort to the latter. Again, this may be attributed to changes in syllabus. Further, we also noticed that the higher level prerequisite CSCI 4061 (Introduction to Operating Systems) does not appear as a causal course. This may be due to the fact that we are dealing with the undergraduate population and not many students enroll in higher graduate level courses (this is seen in our data as these courses have less than or equal to 100 students enrolled). Another possible reason for this could be that students who enroll for CSCI 5103 do so directly without having taken CSCI 4061 in a prior semester.

We observed results for multiple other pairs of courses in a similar fashion and were able to see that our method is successfully able to identify the prerequisite courses as causal. Further, we also observed that the average RMSE for all pairs of $(Y,X)$ courses is fairly low indicating that our regression model is a good fit. Thus, our method may be used in identifying higher level causal courses as well.

\subsection{Sensitivity Analysis}
We test our methodology for robustness by varying a certain hyper-parameter of our model. We vary the distance cut-off during the 1:1 matching, in the range (0.1, 0.3, 0.4, 0.5, 0.6, 0.9). We run our entire analysis for each of these cut-off values and then compute a set-based similarity measure to evaluate how similar our results will be for the different cut-off values. For each distance $d$ in our range above, we do the following: for every valid course $Y$, we obtain the $ATE_{reg}$ for every valid prior course $X$. We then find ${Top3X}(Y, d)$, which are the top 3 causal prior $X$ courses sorted by $ATE_{reg}$. Once we have computed ${Top3X}(Y, d)$ for all combinations of $Y$ and $d$, we compute a similarity measure between the ${Top3X}(Y, d)$ for every pair of distances as follows:
\begin{displaymath}
    Sim(d,d') = \frac{1}{3}\sum_{\forall Y}{{Top3X}(Y,d) \cap {Top3X}(Y,d')},
\end{displaymath}
where, $d$ and $d'$ is every pair of distances from the range given above.
    
We test the robustness of our methodology using both cohorts of data (Table 2). It is shown that for a strict distance cut-off such as 0.1, the Top 3 courses agree with those from higher distances only about 51\% -- 76\% of the time. Upon further investigation it was seen that these matches correspond only to the lower level prerequisite courses. Whereas, for distances 0.3 and 0.4, we see a higher degree of agreement. We also see a high degree of agreement for more relaxed distance cut-offs of 0.6 and above, however, choosing such a relaxed cut-off would mean that we choose students $(s^T, s^C)$ that are very different from each other, thus disrupting the matching between the Treatment and Control group. We also evaluate this robustness check for the Top 5 and Top 1 causal courses for every course $Y$. Overall, we see that our methodology is robust and is slightly sensitive at extreme values of the cut-off distance.          
\balance

\section{Conclusion}
This study aimed to find causal relationships between courses through the methods of causal inference in observational data. We proposed an approach based on the Rubin causal model\cite{Rubin} and used a doubly robust method of matching and regression to obtain estimates of the causal effect (or ATE) of a course $X$ on a subsequent course $Y$. We implemented this methodology on the dataset obtained from the Computer Science department at the University of Minnesota from the years 2002 -- 2016. The results show that this approach may be used to identify courses with causal relationships and can eventually be used to aid in building better academic pathways to help students in completing their degree programs.

\subsection{Acknowledgements}       
This work was supported in part by NSF(1447788, 1704074, 1757916, 1834251), Army Research Office(W911NF1810344), Intel Corp, and the Digital Technology Center at the University of Minnesota. Access to research and computing facilities was provided by the Digital Technology Center and the Minnesota Supercomputing Institute. http://www.msi.umn.edu

%
% The next two lines define the bibliography style to be used, and the bibliography file.
\bibliographystyle{acm}
\bibliography{capstone}

\end{document}